\documentclass[twoside,web]{ieeecolor}
\usepackage{generic}
\usepackage{cite}
\usepackage{amsmath,amssymb,amsfonts}
\usepackage{algorithmic}
\usepackage{graphicx}
\usepackage{textcomp}

\def\BibTeX{{\rm B\kern-.05em{\sc i\kern-.025em b}\kern-.08em
    T\kern-.1667em\lower.7ex\hbox{E}\kern-.125emX}}
\DeclareUnicodeCharacter{03BC}{\ensuremath{\mu}}
\begin{document}

\title{Experiments and Modeling of Defect Dynamics and BTI Behavior in Doped InO TFTs during 400$^\circ$C Post-Processing Forming Gas Annealing}

\author{Yu-Hsin Kuo, Chengyang Zhang, Priyankka Ravikumar, Sanghyun Kang, Taeyoung Song, Marco Villena, Luca Larcher, Hwan Kim, Minji Hong, Pilsang Yun, Gaurav Thareja, Shimeng Yu, Daewon Ha, Suman Datta, Julia Medvedeva and Asif Khan
\thanks{Author accepted manuscript.
Published in IEEE Transactions on Electron Devices,
vol. 73, no. 5, pp. 3140--3148, 2026.
DOI: 10.1109/TED.2026.3658060.}
\thanks{
\copyright~2026 IEEE.  Personal use of this material is permitted.  Permission from IEEE must be obtained for all other uses, in any current or future media, including reprinting/republishing this material for advertising or promotional purposes, creating new collective works, for resale or redistribution to servers or lists, or reuse of any copyrighted component of this work in other works.
}
\thanks{This work was supported by SRC-GRC-NMP program, Samsung Electronics Co., Ltd (IO230407-05815-01), DOE, EERE, SETO program (grant DE-EE0009346), NSF and DARPA MTO’s FLEX program. Fabrication was performed at the IMS, supported by the NSF-NNCI program (ECCS-1542174). The computational resources were provided by NSF-MRI grant OAC-1919789}
\thanks{Y.K, C.Z, P.R, S.K, T.Y, S.Y, S.D and A.K are with the School of ECE, Georgia Institute of Technology, Atlanta, GA, 30318, USA (e-mail: ykuo65@gatech.edu, akhan40@gatech.edu).}
\thanks{M.V is with University of Granada, Spain.}
\thanks{L.L and G.T are with Applied Materials, USA.}
\thanks{H.K, M.H, P.Y and D.H are with Semiconductor Research and Development, Samsung Electronics Co., Ltd., South Korea.}
\thanks{J.M is with Missouri University of Science and Technology, MO, USA.}
\thanks{S.D and A.K are with School of Material Science and Engineering, Georgia Institute of Technology, Atlanta, GA, 30318, USA}
}

\maketitle
\thispagestyle{plain}

\begin{abstract}
We investigate the impact of a monolithic three-dimensional (M3D) integration process-critical 400$^\circ$C post-processing forming gas anneal (FGA) on the electrical performance, reliability, and defect evolution of oxide-channel thin-film transistors (TFTs), combining systematic experiments with density-functional-theory (DFT)–based liquid-quench molecular-dynamics (MD) simulations. Indium tungsten oxide (IWO) TFTs are employed as a model system and encapsulated with a thin 3 nm Al$_2$O$_3$ / 3 nm HfO$_2$ hybrid layer that effectively suppresses external hydrogen ingress. We reveal a non-monotonic evolution of device behavior during FGA, governed by initial densification followed by partial crystallization. Short-duration FGA (10 min) induces channel densification and the formation of shallow, delocalized defect states, leading to pronounced positive bias temperature instability (PBTI) degradation and the emergence of a characteristic transfer-curve “kink.” With prolonged annealing ($>$40 min), partial crystallization of the oxide channel occurs, stabilizing hydrogen in deep, localized defect states, suppressing hydrogen mobility, and restoring device reliability. As a result, the PBTI shift is reduced to 10.4 mV after 2000 seconds of stress, accompanied by complete elimination of the transfer-curve kink. These findings provide a mechanistic understanding of hydrogen–defect interactions during high-temperature post-processing FGA and demonstrate that appropriate hydrogen-blocking encapsulation enables oxide-channel TFT integration without compromising electrical performance or reliability.
\end{abstract}


\begin{figure}[t]
\centerline{\includegraphics[width=\columnwidth]{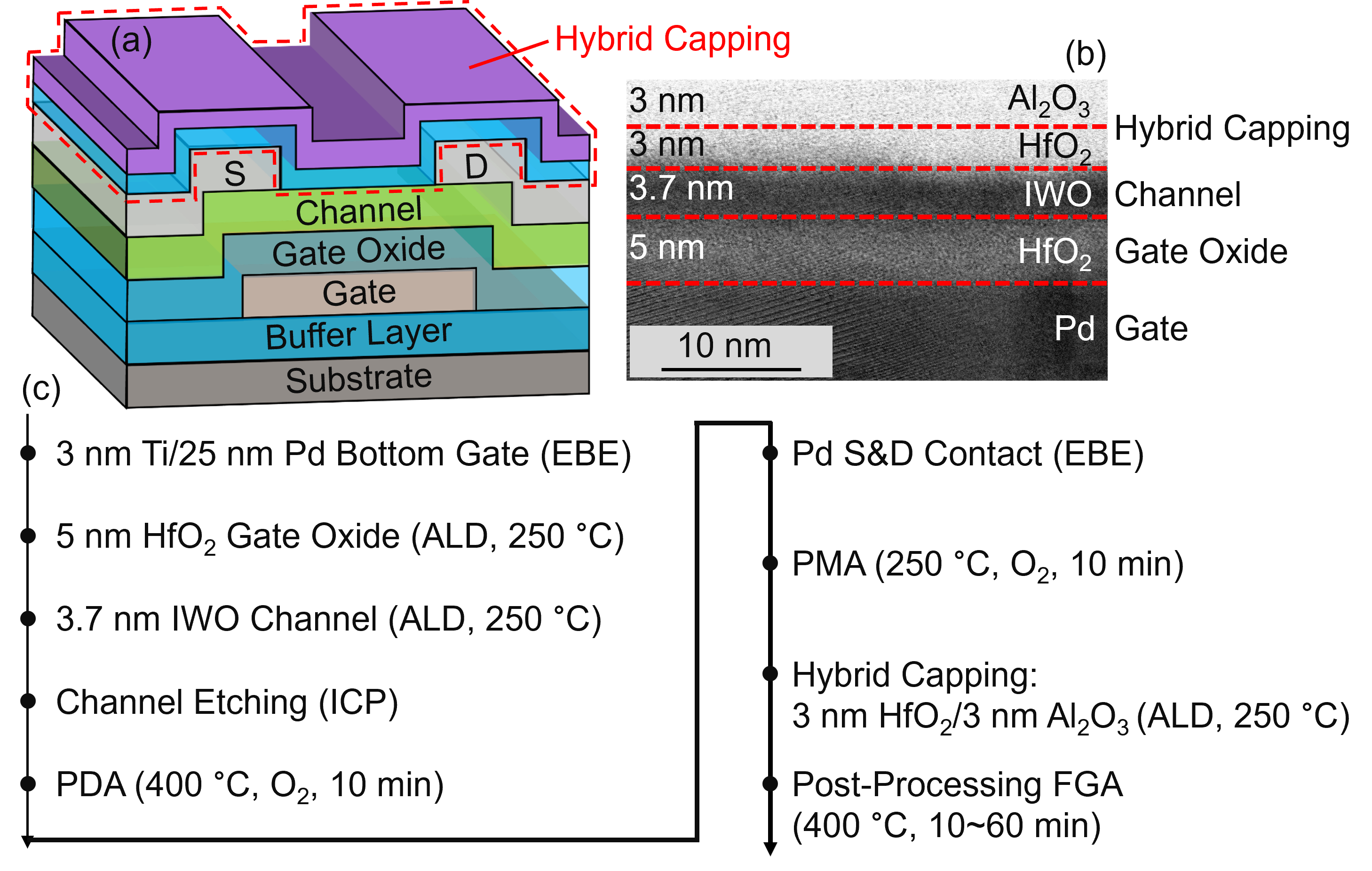}}
\caption{(a) 3D schematic of a BG-LC IWO TFT with optimized hybrid capping layer stack. (b) TEM image of the channel region of an IWO TFT. (c) Key fabrication process flow }
\label{fig1}
\end{figure}

\begin{figure}[t]
\centerline{\includegraphics[width=\columnwidth]{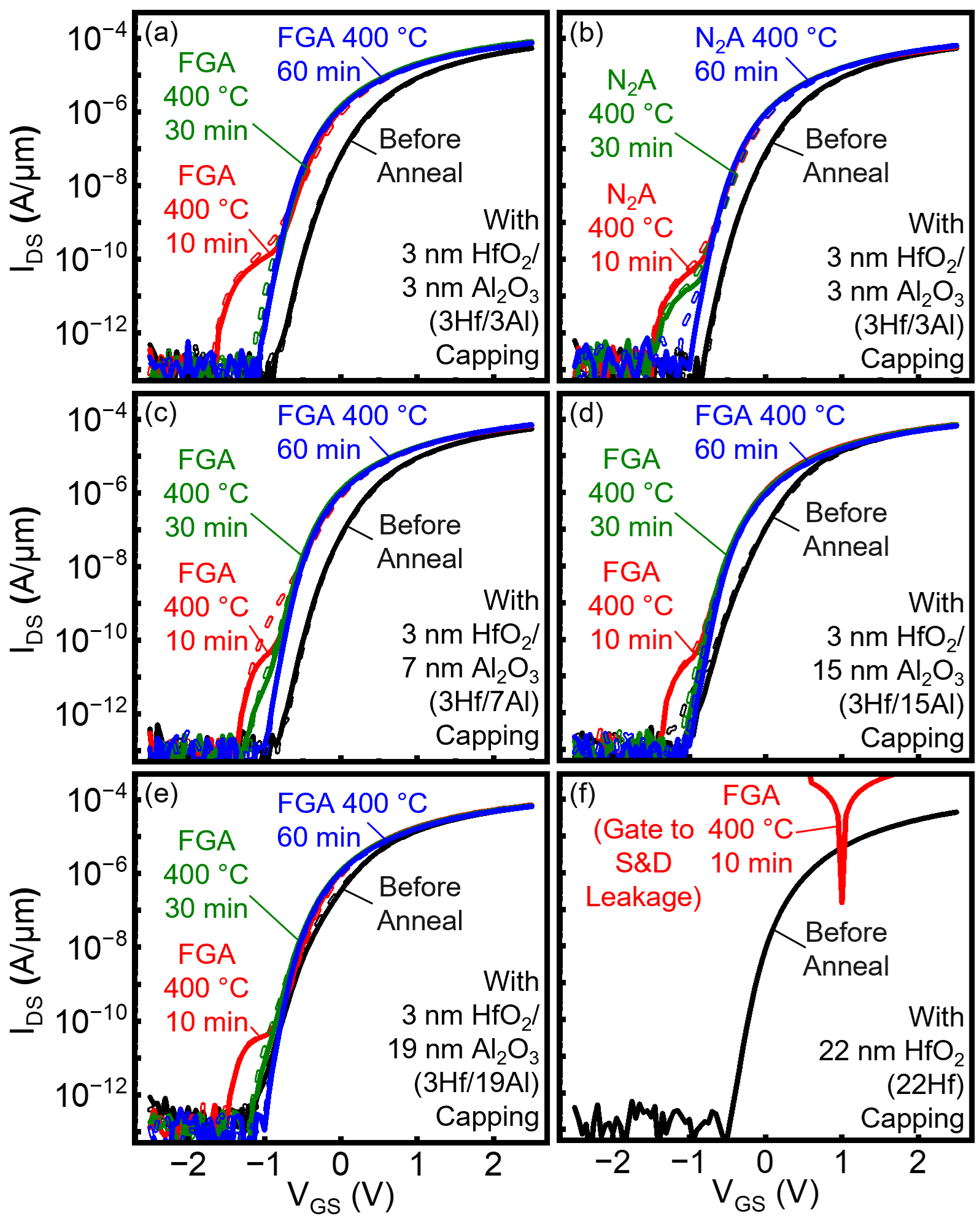}}
\caption{I$_{DS}$-V$_{GS}$ evolution after (a) FGA with a 3 nm HfO$_2$/3 nm Al$_2$O$3$ (3Hf/3Al) capping layer; (b) N$_2$A with a 3 nm HfO$_2$/3 nm Al$_2$O$3$ (3Hf/3Al) capping layer; (c) FGA with a 3 nm HfO$_2$/7 nm Al$_2$O$3$ (3Hf/7Al) capping layer; (d) FGA with a 3 nm HfO$_2$/15 nm Al$_2$O$3$ (3Hf/15Al) capping layer; (e) FGA with a 3 nm HfO$_2$/19 nm Al$_2$O$3$ (3Hf/19Al) capping layer; and (f) FGA with a 22 nm HfO$_2$ (22Hf) capping layer.}
\label{fig2}
\end{figure}

\begin{figure}[t]
\centerline{\includegraphics[width=\columnwidth]{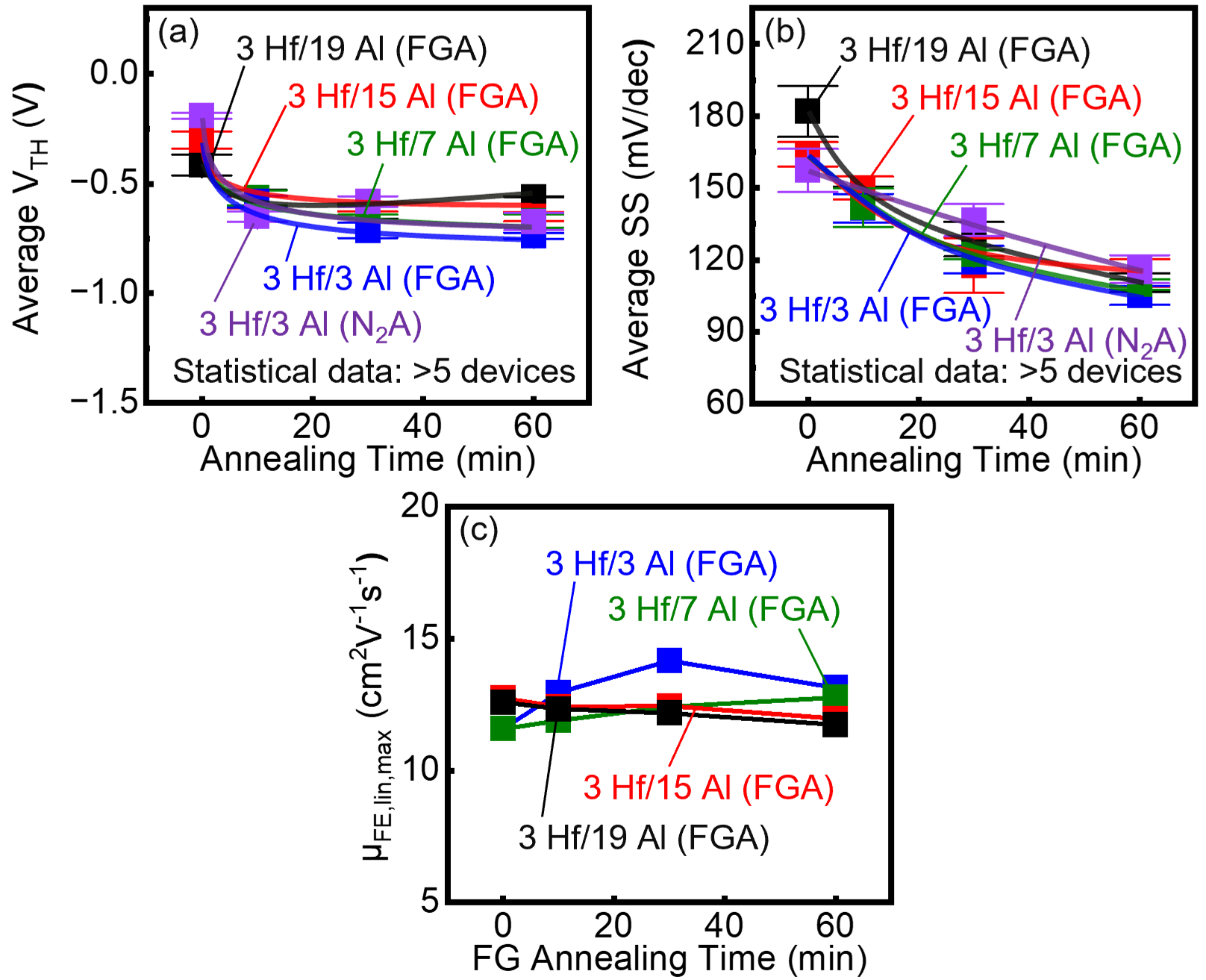}}
\caption{Statistical trends of (a) V$_{TH}$ and (b) SS across different FGA and N$_2$A durations for devices with various hybrid capping layers. (c) Extracted µ$_{FE,lin,max}$ as a function of FGA time for 4 different hybrid capping stacks.}
\label{fig3}
\end{figure}

\section{Introduction}
\label{sec:Introduction}
Indium oxide (InO)–based TFTs are promising candidates for high-performance M3D integration due to their excellent electrical properties and compatibility with back-end-of-line (BEOL) thermal budgets \cite{SDatta2024VLSI, SDeng2023IEDM}. However, a critical and unavoidable step in M3D integration—the FGA—poses major performance, stability, and reliability challenges for oxide-semiconductor (OS)-TFTs. FGA at ~400$^\circ$C is an essential process in advanced complementary metal–oxide–semiconductor (CMOS) manufacturing, where it passivates dangling bonds in front-end silicon logic transistors, improves metal-line reliability, heals plasma-induced damage, and reduces contact resistance \cite{JHLee2023IRPS, WXiong2004EDL}.

Despite these benefits, the hydrogen-rich FGA environment can destabilize OS-TFTs, inducing performance degradation such as threshold-voltage (V$_{TH}$) shifts and subthreshold-swing (SS) degradation through hydrogen incorporation and defect generation \cite{ZLin2025EDL, HTang2025EDL}. Moreover, FGA can exacerbate bias-temperature instability (BTI), as hydrogen diffuses into the oxide channel during bias stress, effectively doping the channel and causing pronounced V$_{TH}$ shifts over time \cite{AChasin2024IEDM}. To address these challenges, prior studies have primarily pursued two strategies: (1) channel engineering, which enhances intrinsic hydrogen tolerance through approaches such as fluorine treatment or dopant modification \cite{BTang2025VLSI, AKruv2025ACSAEM}, and (2) barrier engineering, which employs hydrogen-blocking encapsulation layers \cite{GLiu2025VLSI, Qing2024IEDM}. However, despite these efforts, a direct and comprehensive understanding of how post-processing FGA influences OS-TFT performance and reliability remains lacking. In particular, the evolution of defects during FGA and their direct correlation with electrical degradation have not been systematically elucidated.

In this work, we present a combined experimental and theoretical investigation of capped OS-TFTs subjected to 400$^\circ$C post-processing FGA for up to 60 min, corresponding to the upper BEOL thermal limit. By integrating electrical characterization with technology computer-aided design (TCAD) and DFT–based MD simulations, we directly reveal a two-stage defect evolution during FGA that closely correlates with measured device behavior. These insights provide critical guidance for enabling reliable integration of OS-TFTs in advanced BEOL process flows.

\section{Device Fabrication Details}
\label{sec:Device Fabrication Details}
Fig. 1(a) shows a three-dimensional (3D) schematic of the bottom-gate (BG) long-channel (LC) IWO TFTs investigated in this work. Device fabrication proceeded as follows. First, a Ti/Pd (3 nm/25 nm) bottom gate electrode was deposited by e-beam evaporation (EBE) and patterned using a lift-off process. A 5 nm HfO$_2$ gate dielectric was then deposited by plasma-enhanced atomic layer deposition (PE-ALD) at 250$^\circ$C. Subsequently, a 3.7 nm thick 4 \% W-doped InO (IWO) channel layer was deposited by PE-ALD at 250$^\circ$C. The IWO active region was defined by dry etching, followed by post-deposition annealing (PDA) in an oxygen (O$_2$) ambient at 400$^\circ$C for 10 min. Next, a 40 nm Pd layer was deposited and patterned to form the source/drain (S\&D) contacts, followed by post-metallization annealing (PMA) in an O$_2$ ambient at 250$^\circ$C for 10 min. Finally, a HfO$_2$/Al$_2$O$_3$ hybrid capping layer was deposited by ozone-based ALD process at 250$^\circ$C.

Fig. 1(b) presents a cross-sectional transmission electron microscopy (TEM) image confirming the device stack. Electrical characterization was performed using an Agilent B1500A semiconductor parameter analyzer. All measurements were conducted at a drain voltage (V$_{DS}$) of 1.0 V on devices with a channel length (L$_{ch}$) of 1 µm and a channel width (W$_{ch}$) of 10 µm.

\begin{figure}[t]
\centerline{\includegraphics[width=\columnwidth]{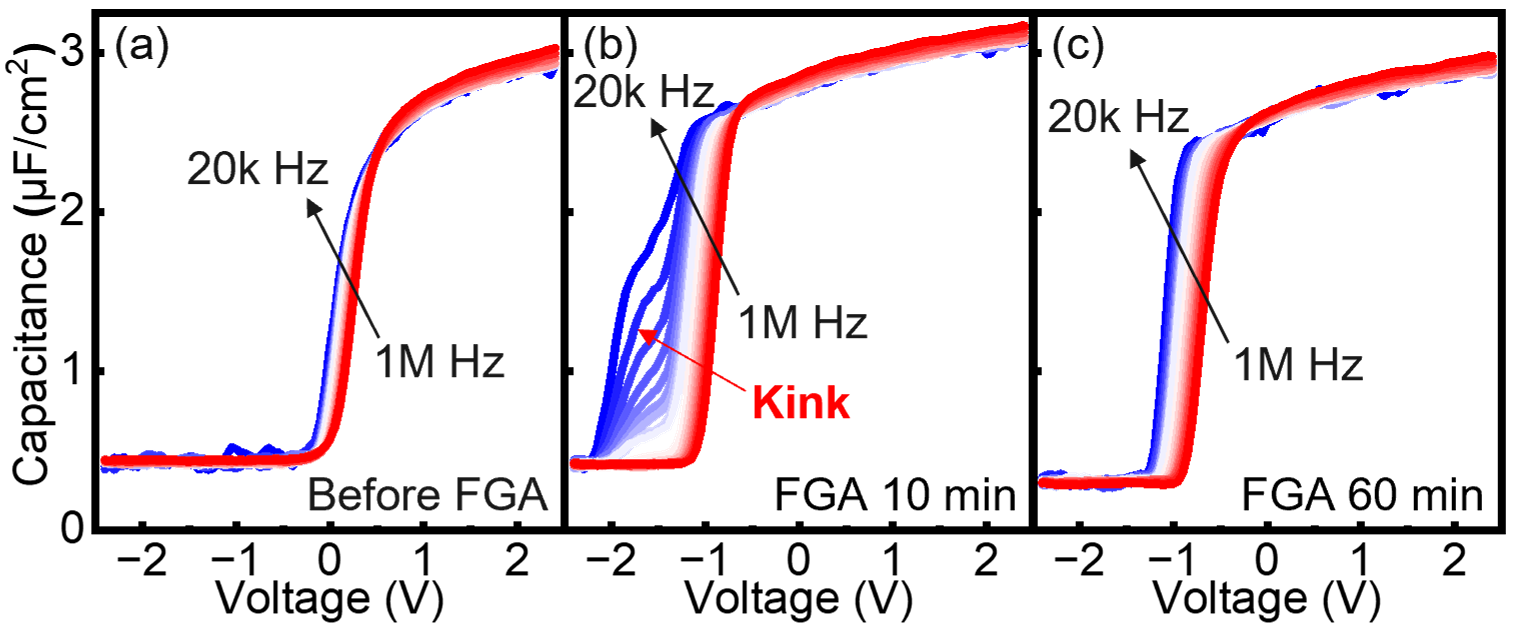}}
\caption{Multi-frequency C–V characteristics under three conditions: (a) before FGA, (b) after 10 min FGA, and (c) after 60 min FGA.}
\label{fig4}
\end{figure}

\begin{figure*}[t]
\centering
  \includegraphics[width=\textwidth]{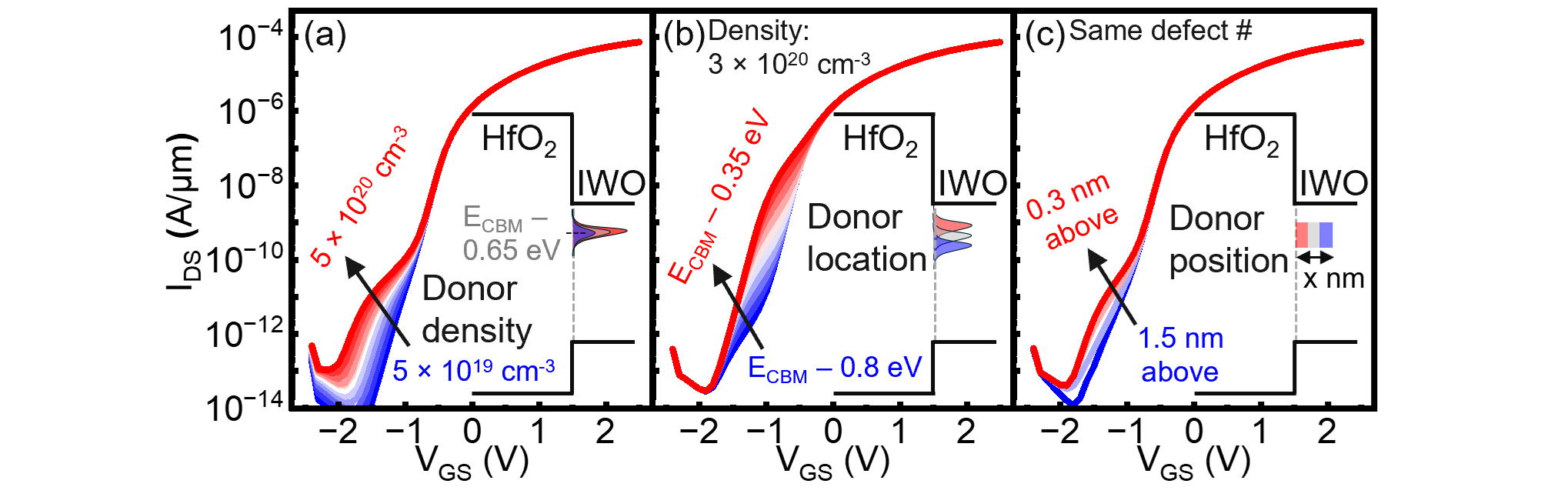}
  \caption{TCAD simulations using the Ginestra™ platform showing I$_{DS}$-V$_{GS}$ characteristics with Gaussian donor states varied in donor state (a) density, (b) energy level, and (c) spatial location.}
\label{fig5}
\end{figure*}

\section{Hydrogen Immunity toward 400$^\circ$C FGA}
\label{sec:Hydrogen Immunity toward 400C FGA}
Fig. 2(a) shows the evolution of transfer characteristics (I$_{DS}$–V$_{GS}$) of IWO TFTs capped with an optimized 6 nm hybrid stack consisting of 3 nm HfO$_2$/3 nm Al$_2$O$_3$ (3Hf/3Al) and subjected to 400$^\circ$C post-processing FGA for durations from 10 to 60 min. A slight V$_{TH}$ shift is observed after the first 10 min of FGA, after which V$_{TH}$ remains stable with prolonged annealing up to 60 min. The HfO$_2$ interlayer is critical for suppressing post-capping degradation, as direct Al$_2$O$_3$ deposition was found to induce V$_{TH}$ shifts, likely due to hydrogen-related by-products generated during the Al$_2$O$_3$ ALD deposition process. However, HfO$_2$ alone provides insufficient hydrogen-blocking capability, leading to significant gate-to-source/drain leakage after 10 min of FGA, as shown in Fig. 2(f), rendering the device nonfunctional. In contrast, Al$_2$O$_3$ exhibits excellent hydrogen-blocking capability \cite{HKKo2007ISAF,YKKim1997IEDM}. As shown in Fig. 2(e), replacing the top 19 nm HfO$_2$ layer with Al$_2$O$_3$ results in stable V$_{TH}$ during 400$^\circ$C FGA for up to 60 min. Moreover, the optimized Al$_2$O$_3$ thickness can be scaled down to 3 nm, as demonstrated in Fig. 2(a) and Fig. 2(c–e), without compromising hydrogen-blocking effectiveness.

To isolate the role of hydrogen during annealing, a control experiment using pure nitrogen annealing (N$_2$A) at 400$^\circ$C with the same capping stack was performed, as shown in Fig. 2(b). Fig. 3(a-b) summarize the statistical trends of V$_{TH}$ and subthreshold swing (SS) as a function of Al$_2$O$_3$ capping thickness under FGA, together with the N$_2$A control. Consistent trends are observed across all conditions, indicating that the observed performance evolution is primarily thermally driven rather than ambient hydrogen ingression. V$_{TH}$ is extracted at a constant I$_{DS}$ of 10 nA/µm, and SS is defined as the inverse slope of the transfer curve one decade I$_{DS}$ change below V$_{TH}$. Owing to effective hydrogen blocking, a substantial improvement (approximately 30\%) in SS is achieved after 60 min of FGA using optimized 3Hf/3Al capping layer, indicating a reduced interface trap density.

Furthermore, we extracted the maximum linear field-effect mobility (µ$_{FE,lin,max}$) as a function of FGA time for devices with different hybrid capping stacks to evaluate whether carrier scattering changes with prolonged FGA. As shown in Fig. 3(c), µ$_{FE,lin,max}$ remains essentially independent of FGA time up to 60 min for all capping stacks. The mobility was extracted using:
\begin{equation}
\mu_{\mathrm{FE,max,lin}} = g_{m,\mathrm{max}}\left(\frac{L_{\mathrm{ch}}}{W_{\mathrm{ch}}\,C_{\mathrm{ox}}\,V_{\mathrm{DS}}}\right),\quad
g_m=\frac{\partial I_{\mathrm{DS}}}{\partial V_{\mathrm{GS}}}.
\label{eq:mufe}
\end{equation}
Here L$_{ch}$ = 1 µm, W$_{ch}$ = 10 µm, C$_{ox}$ (gate oxide capacitance) = 2.94 µF/cm$^2$, V$_{DS}$ = 1.0 V and g$_{m,max}$ is the maximum transconductance over the measured V$_{GS}$ sweep range. C$_{ox}$ was experimentally extracted from capacitance–voltage (C–V) measurements performed on the TFT structure. This suggests that prolonged FGA does not significantly affect carrier scattering in our devices, which is beneficial for BEOL integration without degrading performance.

\begin{figure*}[t]
\centering
  \includegraphics[width=\textwidth]{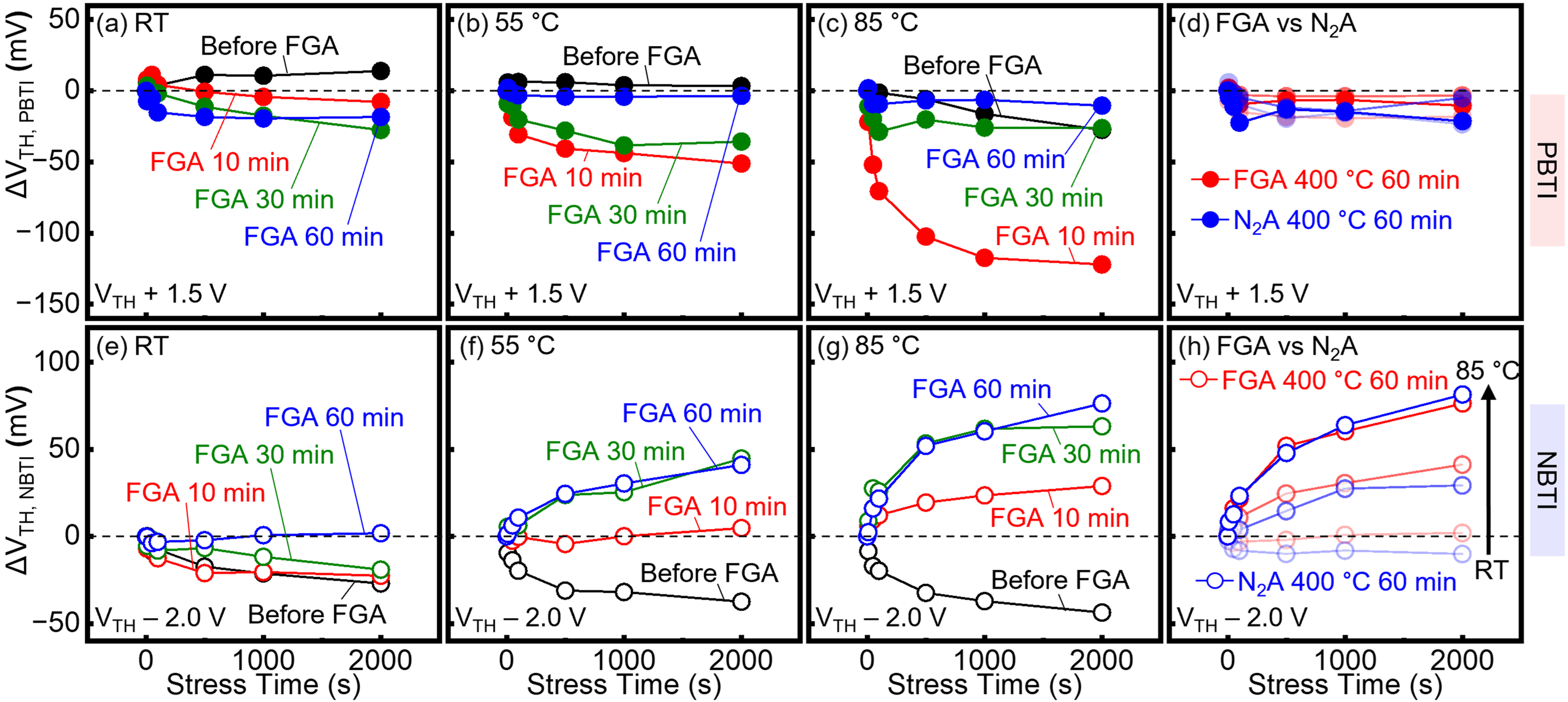}
  \caption{BTI evolution under post-processing FGA and N$_2$A. PBTI (a–c) and NBTI (e–g) characteristics of IWO TFTs under four conditions: (1) before FGA, and after 400$^\circ$C FGA for (2) 10 min, (3) 30 min, and (4) 60 min. Panels (d) and (h) compare the effects of FGA and N$_2$A treatments.}
\label{fig6}
\end{figure*}

\begin{figure}[t]
\centerline{\includegraphics[width=\columnwidth]{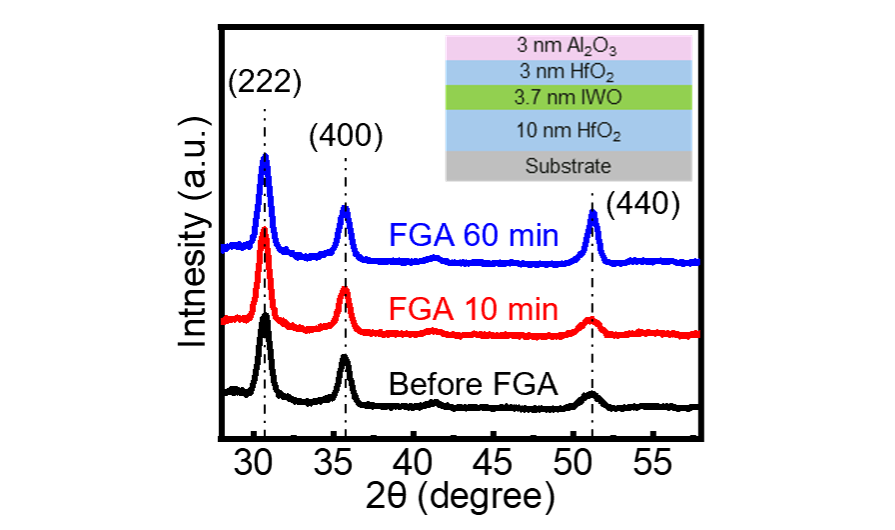}}
\caption{GI-XRD analysis of a 3.7 nm IWO film with an optimized hybrid capping layer (3Hf/3Al) under three conditions: (1) before FGA, and after FGA for (2) 10 min and (3) 60 min. A pronounced increase in the intensity of the (440) diffraction peak is observed after 60 min of FGA, indicating partial crystallization of the IWO channel.}
\label{fig7}
\end{figure}

\section{“Kink” Formation during Post-Processing FGA}
\label{sec:“Kink” Formation during Post-Processing FGA}
A distinct “kink” feature emerges in the I$_{DS}$–V$_{GS}$ characteristics in the deep subthreshold regime (approximately 10 pA/µm) after 10 min of both post-processing FGA and N$_2$A. This anomalous feature gradually diminishes with prolonged annealing, as shown in Fig. 2(a-b). Consistently, multi-frequency C–V measurements in Fig. 4(a–c) reveal the same "kink" behavior after 10 min, which diminishes after 60 min, indicating a transient effect associated with intermediate defect states that form during early annealing and stabilize with extended annealing.

To elucidate the physical origin of this behavior, numerical simulations were performed using the Ginestra™ platform, showing excellent agreement with the experimental observations. As shown in Fig. 5(a–c), introducing a Gaussian distribution of shallow donor states near the conduction band minimum (CBM) of IWO and spatially localized close to the gate-oxide/IWO interface successfully reproduces the observed "kink" feature. These results confirm that the 10-min annealing–induced "kink" originates from an increased population of shallow interface donor states. The "kink" behavior is found to be sensitive to three key parameters: (1) defect density, where the donor energy level is fixed at 0.65 eV below the CBM and positioned 0.3 nm above the gate-oxide/IWO interface, as shown in Fig. 5(a); (2) defect energy level, where the donor density is fixed at 3e20 cm$^{-3}$ and positioned 0.3 nm above the interface, as shown in Fig. 5(b); and (3) spatial location, where both the donor energy level (0.65 eV below the CBM) and total donor concentration are fixed, as shown in Fig. 5(c).

These shallow donor states enhance trap-assisted transport (TAT) \cite{WChakraborty2020VLSI}, particularly in the deep subthreshold regime, where carriers can be thermally activated from trap states into extended states above the CBM. Here, we use TAT in a broad sense to denote trap-mediated conduction, including (1) tunneling/hopping via localized trap states and (2) dynamic trapping/detrapping during the direct-current (DC) gate-voltage sweep enabled by densification-driven structural reconfiguration. Specifically, the 3- to 4-coordinated oxygen transformation shifts some deep trap states toward the CBM, creating weakly localized states \cite{JMedvedeva2022PRM,JMedvedeva2020JAP}. Thermal activation of these states leads to premature channel turn-on and the characteristic transfer-curve “kink”.

\section{Evolution of BTI under post-processing FGA}
\label{sec:Evolution of BTI under post-processing FGA}
A comprehensive reliability study was conducted across different FGA durations at operating temperatures of room temperature (RT), 55$^\circ$C, and 85$^\circ$C under a fixed gate overdrive (V$_{OV}$) stress. DC bias stressing was applied, followed by an immediate fast current–voltage readout (100 µs) to accurately capture degradation while minimizing recovery effects.

Fig. 6(e–g) show the evolution of negative BTI (NBTI) for four post-processing conditions: (1) before FGA, and after 400$^\circ$C FGA for (2) 10 min, (3) 30 min, and (4) 60 min, measured at different operating temperatures. Before FGA, the devices exhibit a negative threshold voltage shift ($\Delta$V$_{TH}$), originating from field-induced ionization of oxygen vacancies in the oxide channel \cite{KAAabrar2024VLSI}. During negative bias stress, these vacancies release electrons, increasing channel conductivity and resulting in a negative $\Delta$V$_{TH}$.

After FGA, a polarity reversal of the $\Delta$V$_{TH}$ is observed, changing from negative to positive, with the effect becoming more pronounced at higher operation temperatures and longer FGA durations. This reversal is attributed to diffusion of intrinsic hydrogen. Because the devices are immune to hydrogen from the annealing ambient, as demonstrated in the previous discussion, the observed intrinsic hydrogen is attributed to the Pd S\&D contacts, which can absorb and release hydrogen during fabrication due to their strong hydrogen catalytic activity \cite{YShi2024ACSnano}. Under electrical stress, diffusion of this intrinsic hydrogen contributes significantly to the observed instability.

Fig. 6(a–c) present the corresponding positive BTI (PBTI) results. A similar polarity reversal of the $\Delta$V$_{TH}$ is observed after FGA, from positive to negative, again linked to intrinsic hydrogen diffusion. Notably, after 10 min of FGA, a strong negative $\Delta$V$_{TH}$ appears, which gradually weakens with longer annealing durations. This trend is attributed to partial crystallization of the oxide channel during prolonged FGA. As shown by the grazing-incidence X-ray diffraction (GI-XRD) results in Fig. 7, a 60 min FGA increases the crystallinity of the IWO channel film, as evidenced by the pronounced enhancement of the (440) diffraction peak intensity. As crystallization proceeds, hydrogen preferentially forms strong oxygen–hydrogen (O-H) bonds at crystalline–amorphous (c/a) interfaces, effectively immobilizing hydrogen and suppressing its diffusion \cite{JMedvedeva2022ACSAMI}. Consequently, the magnitude of the negative $\Delta$V$_{TH}$ decreases with extended annealing time. These experimental observations will be revisited in conjunction with DFT modeling in the following section, where a comprehensive BTI mechanism is established.

To further confirm that this effect does not originate from ambient hydrogen, BTI measurements were also performed on control devices annealed in a pure nitrogen ambient at 400$^\circ$C as shown in Fig. 6(d,h). After 60 min of N$_2$A, both PBTI and NBTI exhibit trends similar to those observed after FGA, confirming that the behavior arises from intrinsic hydrogen rather than the annealing ambient hydrogen.

\begin{figure}[t]
\centerline{\includegraphics[width=\columnwidth]{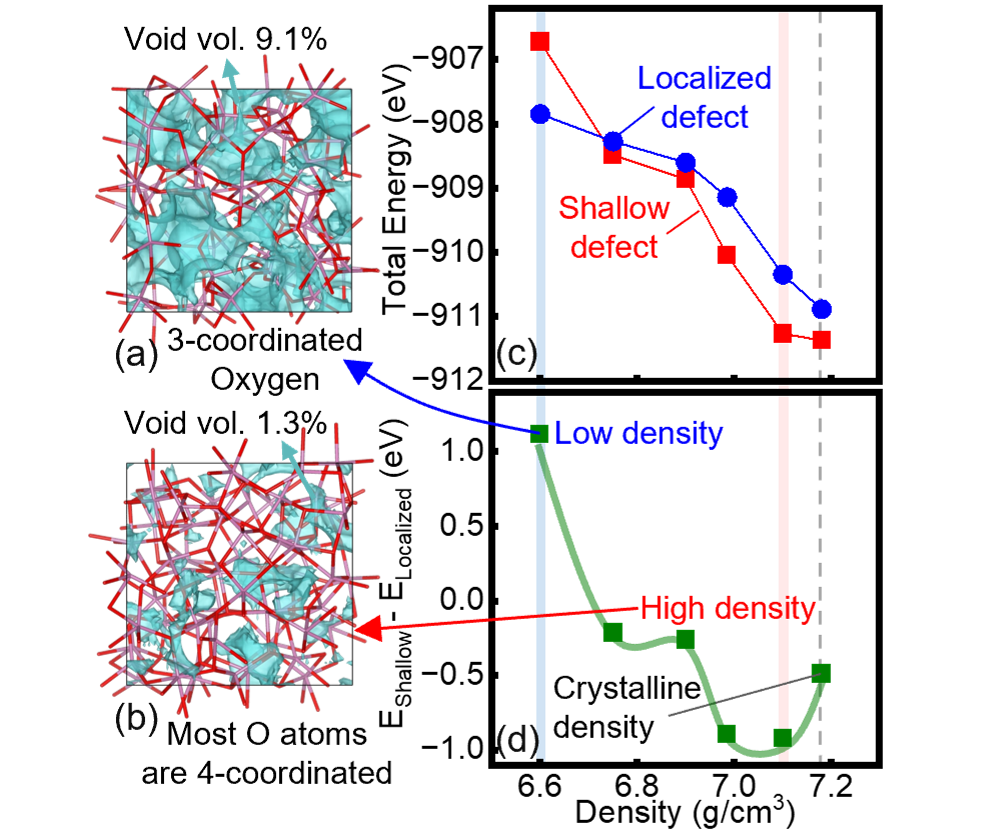}}
\caption{Atomic structures of (a) low- and (b) high-density a-InO highlight differences in void space and coordination. Hybrid-functional calculations in (c) and (d) show the total energy and energy difference of shallow vs. localized defects as a function of density. }
\label{fig8}
\end{figure}

\begin{figure}[t]
\centerline{\includegraphics[width=\columnwidth]{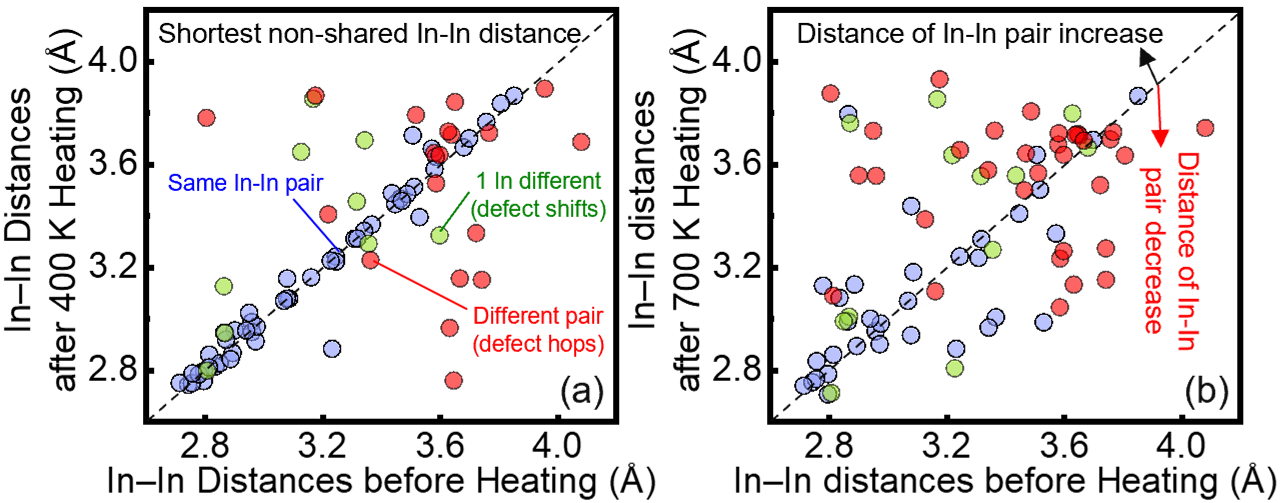}}
\caption{Shortest non-shared In–In distances before and after heating at (a) 400 K and (b) 700 K for a-InO. Color coding indicates defect stability: blue denotes a stable In–In defect that remains unchanged at the given temperature; green indicates an unstable defect that relaxes into a deeper local minimum involving one different In atom while retaining the other; red denotes annihilation of the initial defect, where the original In atoms increase their coordination and no longer localize charge, accompanied by the formation of a new defect involving a different pair of under-coordinated In atoms that trap charge. The newly formed defect may appear either near or far from the original defect site within the supercell.}
\label{fig9}
\end{figure}

\begin{figure}[t]
\centerline{\includegraphics[width=\columnwidth]{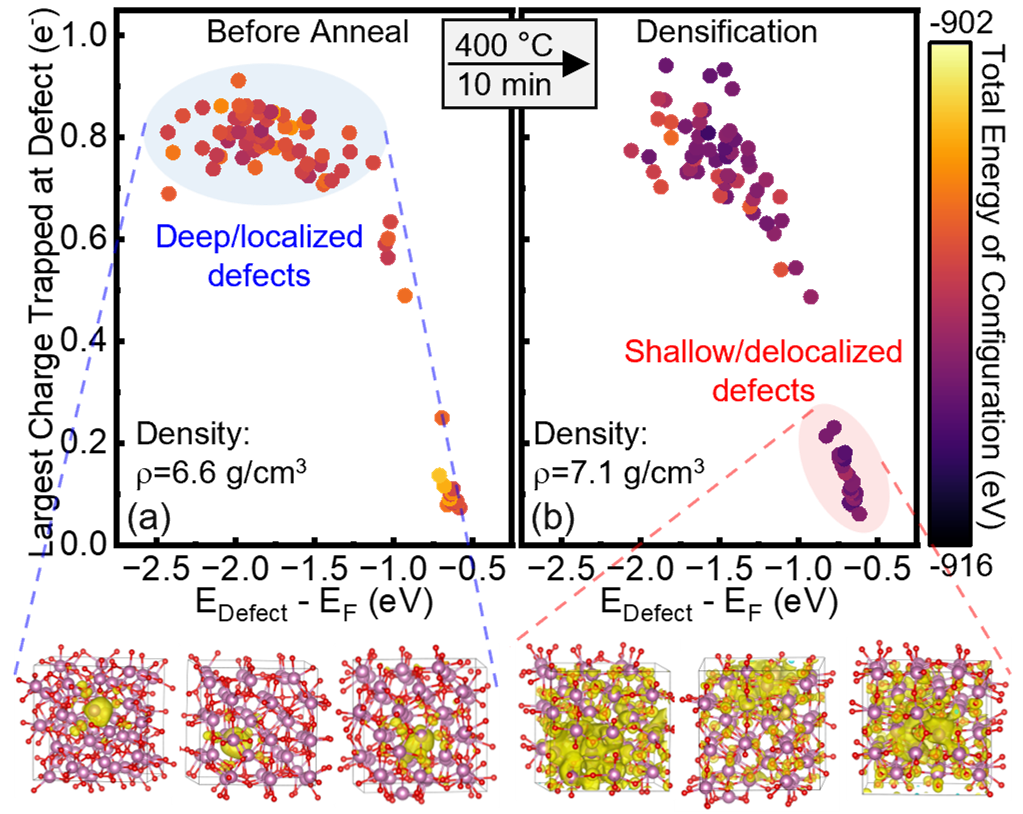}}
\caption{Statistical analysis of 80 independent MD simulations at two a-InO densities: (a) 6.6 g/cm$^{3}$ and (b) 7.1 g/cm$^{3}$. Bottom schematics illustrate representative deep and shallow defect states.}
\label{fig10}
\end{figure}

\section{DFT Simulations of amorphous-InO}
\label{sec:DFT Simulations of amorphous-InO}
DFT calculations provide microscopic insight into the mechanisms underlying the observed behavior of FGA-treated IWO TFTs. All computational models were generated using ab initio molecular dynamics (AIMD) as implemented in the Vienna Ab Initio Simulation Package (VASP) \cite{GKresse1996PRB}. Hydrogen-free and hydrogen-containing amorphous In$_2$O$_{3-x}$ bulk models, as well as c/a In$_2$O$_{3-x}$ interface models, were constructed within DFT using periodic boundary conditions. The Perdew–Burke–Ernzerhof (PBE) exchange–correlation functional and the projector augmented-wave (PAW) method were employed.

To generate amorphous structures, liquid–quench AIMD simulations were performed with quench rates of 50–200 K/ps. During the melting and quenching stages, an energy cutoff of 260 eV and $\Gamma$-point sampling were used. Each quenched structure was subsequently equilibrated at 300 K for 6 ps with an increased cutoff energy of 400 eV. All AIMD simulations were carried out in the NVT ensemble using a Nosé–Hoover thermostat with an integration time step of 2 fs; for H-containing structures, a smaller time step of 0.5 fs was used.

After ~30 ps of AIMD equilibration, the atomic structures were relaxed at 0 K within DFT (PBE). Geometry optimization was performed with a plane-wave cutoff of 500 eV and a $\Gamma$-centered 6×6×6 k-point mesh (36 k-points), until the Hellmann–Feynman forces on all atoms were below 0.01 eV/Å. Charged-supercell calculations were additionally carried out to simulate electron addition/removal, and all charged configurations were fully relaxed. Electronic properties of the PBE-relaxed models were then evaluated using the self-consistent hybrid functional HSE06 \cite{JHeyd2003JCP} with a mixing parameter of 0.25 and a screening parameter $\alpha$=0.2 Å$^{-1}$. Atomic structures and charge-density distributions were visualized using VESTA \cite{KMomma2011JAC}.

Substoichiometric amorphous In$_2$O$_{2.93}$ was modeled using a nonstoichiometric quench approach, which provides improved statistical representation of amorphous structural morphology compared with introducing vacancies into a single amorphous configuration and captures the formation of both shallow states and localized deep defects \cite{JMedvedeva2022PRM}. The c/a interface model was constructed by joining an 80-atom crystalline (bixbyite) In$_2$O$_{3-x}$ slab with an 80-atom amorphous In$_2$O$_{3-x}$ structure, following our prior work \cite{JMedvedeva2022ACSAMI}. This c/a interfacial model was used to investigate defect formation and spatial distributions in the crystalline and amorphous regions as well as at their interface. Overall, our DFT results suggest two distinct stages of defect dynamics during the 60 min FGA: an initial film densification stage occurring at approximately 10 min, followed by partial crystallization during prolonged annealing.

\begin{figure}[t]
\centerline{\includegraphics[width=\columnwidth]{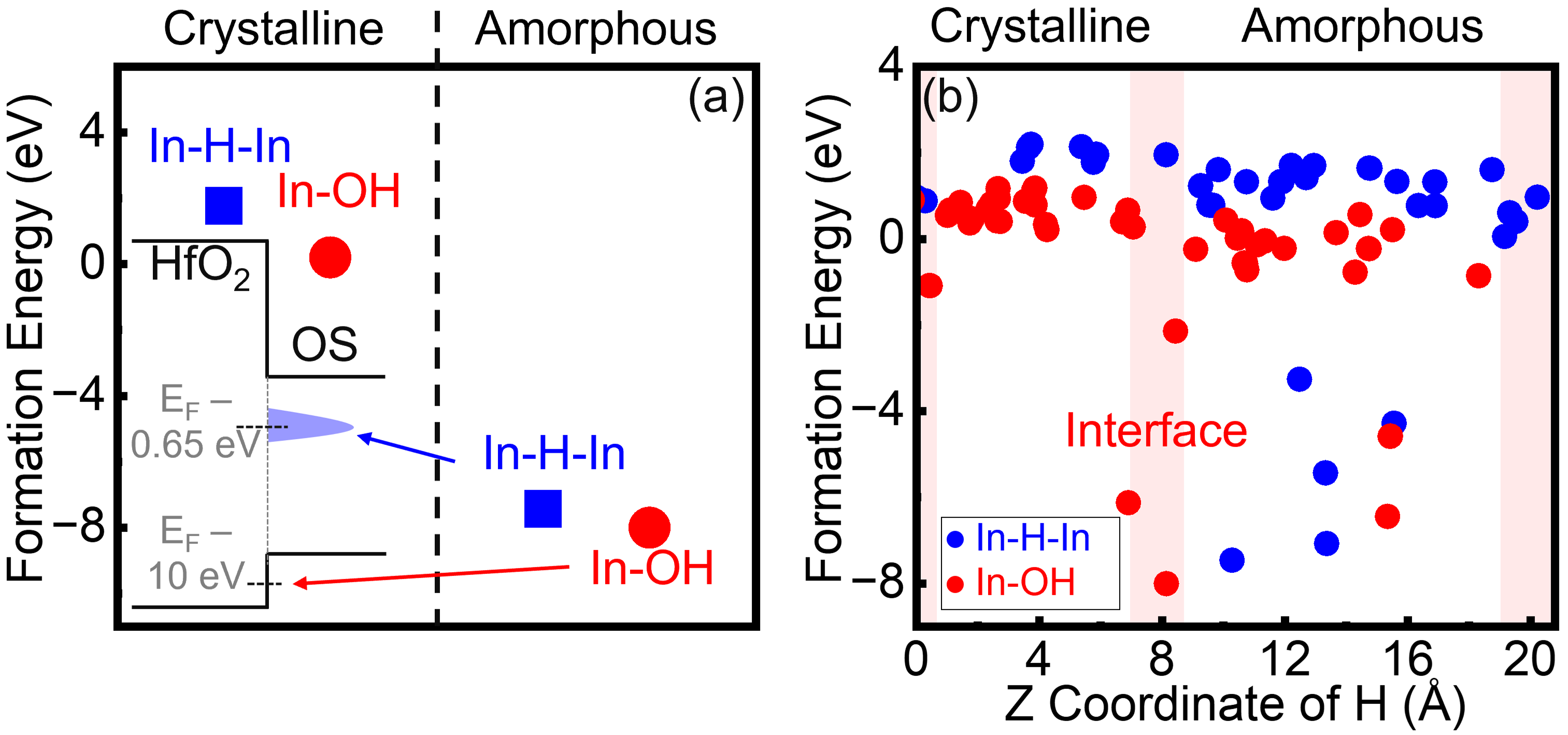}}
\caption{(a) Comparison of the formation energies of covalent In–OH and ionic In–H–In configurations located in crystalline and amorphous regions. (b) Spatial distribution of formation energies for In–H–In and In–OH defects in crystalline regions, amorphous regions, and at crystalline–amorphous interfaces.}
\label{fig11}
\end{figure}

\begin{figure*}[t]
\centering
  \includegraphics[width=\textwidth]{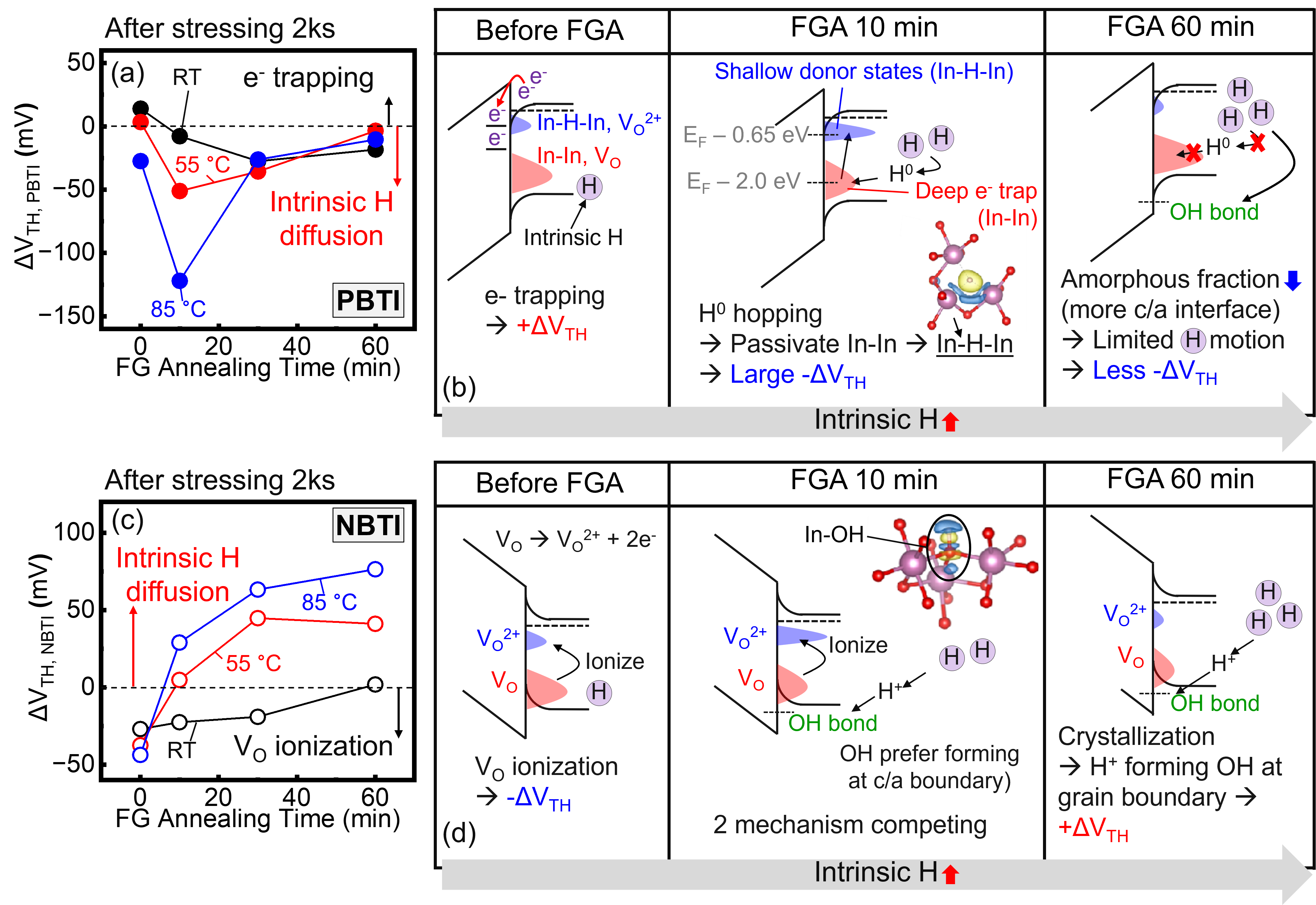}
  \caption{$\Delta$V$_{TH}$ after 2 ks stress is summarized for (a) PBTI and (c) NBTI. Panels (b) and (d) provide a comprehensive summary of the PBTI and NBTI mechanisms, respectively, under different post-processing FGA durations.}
\label{fig12}
\end{figure*}

\subsection{Stage I: Densification in a-InO }
Fig. 8(c-d) present DFT and MD results linking defect energetics to six distinct structural density values in amorphous InO (a-InO) configurations. Low-density structures (6.6 g/cm$^{3}$) exhibit pronounced structural voids and a high fraction of three-coordinated oxygen atoms, as shown in Fig. 8(a). These features give rise to highly localized, deep defect states, reflected by the lower localized defect-state energies observed in Fig. 8(c-d).

MD simulations further indicate that zero-pressure structural relaxation alone is insufficient to eliminate these voids. Instead, the amorphous network must be thermally activated above 350$^\circ$C to transform into a more uniform, void-free polyhedral network dominated by fourfold-coordinated oxygen, characteristic of bixbyite indium oxide, as shown in Fig. 8(b) and Fig. 9(a-b). This theoretically predicted transition temperature is in close agreement with the 400$^\circ$C annealing conditions employed experimentally. Increased oxygen coordination promotes a more uniform polyhedral network that favors the formation of slightly under-coordinated In atoms that act as shallow donor states \cite{JMedvedeva2022PRM}. As shown in Fig. 8(d), in higher-density networks (7.1 g/cm$^{3}$), the energies of these shallow defects are reduced, thereby enhancing electron transport.

A comprehensive analysis of the calculated void distributions and indium and oxygen coordination statistics provides further insight into this density-dependent defect formation. The structural characteristics of defects are correlated with both the amount of charge trapped and the energetic position of defect states relative to the band edges. Fig. 10(a-b) summarize statistical results from 80 independent MD simulations at two representative a-InO densities, offering a clearer picture of how density governs defect behavior. Hybrid-functional electronic-structure and charge-density calculations reveal that strongly localized, deep defects are energetically favored at low density (6.6 g/cm$^{3}$), whereas a larger fraction of defects become shallower at higher density, accompanied by reduced formation energies. Shallow and localized defects differ in (1) the amount of charge trapped at the defect and (2) the energetic position of the defect state relative to the Fermi level. In this work, we define them as follows: (i) Shallow states trap $<$ 0.3 e and their defect levels lie within 1 eV of the Fermi level. (ii) Localized states trap 0.6–1.0 e and their defect levels lie $\sim$1.0–2.5 eV from the Fermi level.

These results indicate that thermally induced densification of as-deposited a-InO is not only energetically favorable but also leads to an increased free carrier concentration. This finding is fully consistent with experimental observations, where shallow donor states emerge after the first 10 min of FGA, giving rise to the characteristic “kink” behavior in the electrical characteristics.

Notably, our simulations span the density range commonly reported for amorphous indium oxide (6.6–7.2 g/cm$^{3}$). Liquid-quench MD for amorphous In–W–O (In$_{50}$W$_4$O$_{87}$) yields an optimal (minimum-energy) density of 6.95 g/cm$^{3}$, only slightly lower than that of W-free amorphous In–O (a-InO, $\sim$7.0 g/cm$^{3}$).

\begin{figure}
\centerline{\includegraphics[width=\columnwidth]{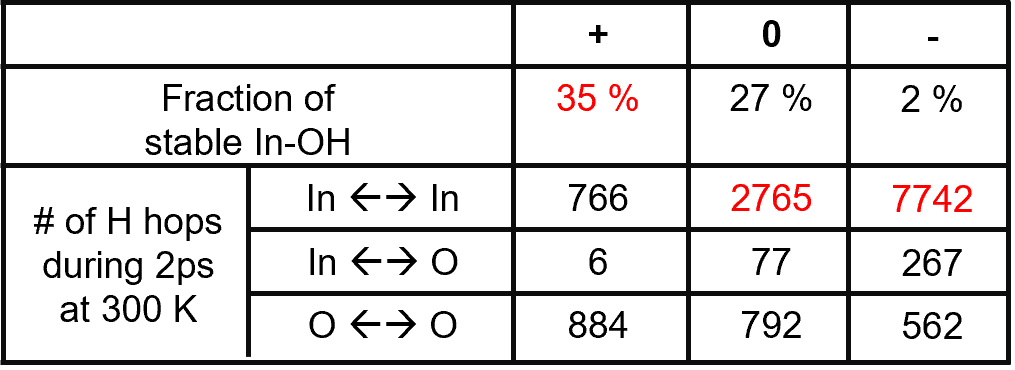}}
\caption{Three charge states ("+", "0" and "-") introduced as a background charge to simulate the corresponding hydrogen charge state were modeled in a-InO, and their behavior was tracked during simulations in terms of: (1) the fraction forming stable In–OH bonds, and the number of hydrogen hops within 2 ps at RT between (2) In–In, (3) In–O, and (4) O–O sites.}
\label{fig13}
\end{figure}

\subsection{Stage II: Partial Crystallization}
Prolonged FGA (60 min) enhances partial crystallinity in the IWO channel film as shown in Fig. 7. In addition, hydrogen is known to play a key role in defect passivation, doping, crystallization, and the enhancement of optical transparency and electrical transport in indium-based oxide semiconductors \cite{JMedvedeva2022ACSAMI}. Consistent with these observations, our DFT results indicate that the intrinsic hydrogen discussed previously preferentially localizes near crystalline–amorphous interfaces.

Hydrogen (H) can bond in two distinct configurations. First, it can passivate metallic In–In pairs by forming ionic In–H–In complexes, which act as shallow donor states. Second, hydrogen can form strong covalent O–H bonds, leaving adjacent highly undercoordinated In–In pairs that behave as deep, localized electron traps. When H transforms from an In–H–In configuration to an In–OH complex, the H motion not only leaves behind an unpassivated In–In pair, but also reduces the In coordination in the In–OH complex because the strong O–H bond weakens the local In–O bonding. The newly created undercoordinated In site may pair with a pre-existing isolated undercoordinated In, forming an additional unpassivated In–In trap. Passivation of these newly created In–In pairs requires additional unbound H, whereas H in the thermally stable In–OH bond is not readily available for re-passivation. As shown in Fig. 11(a), both H-related defect configurations exhibit lower formation energies in amorphous regions, indicating that hydrogen incorporation is energetically favored in the amorphous phase. The formation energy was calculated from the total energy obtained for the fully relaxed structures according to the following equation:
\begin{equation}
E_f(\mathrm{H})_i = E(\mathrm{In}_{64}\mathrm{O}_{96}\mathrm{H}_1)_i - E(\mathrm{In}_{64}\mathrm{O}_{96}) - \frac{1}{2}E(\mathrm{H}_2).
\label{eq:EfH}
\end{equation}
Here, E(In$_{64}$O$_{96}$H$_1$ )$_i$ is the total energy of the H-doped c/a supercell with H at the i-th site (for the specified charge state); E(In$_{64}$O$_{96}$) is the total energy of the pristine c/a supercell (including the corresponding background charge for consistency); and 1/2*E(H$_2$) is half of the total energy of an H$_2$ molecule, taken as the hydrogen chemical potential.
Furthermore, Fig. 11(b) shows that In–OH complexes preferentially form at c/a interfaces, suggesting that hydrogen becomes trapped and immobilized at grain boundaries once partial crystallization occurs.

During grain growth, mobile In–H–In donor configurations are progressively converted into energetically stable In–OH complexes pinned at c/a interfaces. This transformation suppresses H mobility and limits further H-induced doping, providing a microscopic explanation for the recovery of electrical stability observed after prolonged FGA.

\subsection{Role of W Dopant in a-InO}
Regarding the role of W dopants, we performed DFT calculations on mixed c/a In$_2$O$_{3-x}$	models with 40 W-substitution configurations (W on an In site in the crystalline region, amorphous region, or at the c/a interface). The results show that W preferentially resides in the amorphous region: the average substitution energy is $\sim$2.2 eV lower than in the crystalline region and $\sim$1.8 eV lower than at the interface. This suggests that during annealing, W-free nanocrystallites nucleate first, while W segregates to the disordered (amorphous) matrix.

W is not expected to directly participate in H bonding, since formation of W–OH or W–H–W defects is energetically unfavorable due to the strong W–O bond. Instead, W slightly modifies the local InO network: liquid-quench MD (In$_{50}$W$_4$O$_{87}$, 6.9 g/cm$^3$) shows a small reduction in average InO coordination from 5.2 (a-InO) to 5.0 (a-IWO). Consistent with our calculated H hopping parameters (Fig. 13), this reduced coordination correlates with a slightly lower formation energy for In–H–In and increased H mobility.

Finally, the DFT results indicate that crystallization can generate clusters of 3–4 undercoordinated In atoms adjacent to fully coordinated W, often pinned near the c/a interface. These complexes can impede grain growth unless mobile H passivates the undercoordinated In cluster by forming In–H–In.

\section{Comprehensive BTI Mechanisms During FGA}
\label{sec:Comprehensive BTI Mechanisms During FGA}
By combining the electrical results with DFT modeling insights, we establish a comprehensive BTI mechanism across different FGA durations. Fig. 12(a) summarizes the PBTI trend, showing the $\Delta$V$_{TH}$ after 2000 seconds of stress as a function of FGA duration. Before FGA, only a small positive $\Delta$V$_{TH}$ is observed, indicating limited charge trapping. As illustrated in Fig. 12(b), electron trapping at the channel–dielectric interface reduces the effective gate voltage, resulting in a positive $\Delta$V$_{TH}$. After 10 min of FGA, the concentration of intrinsic H increases. The effective-charge-state stability of H is Fermi-level (E$_{F}$) dependent: H$^{-}$ and H$^{0}$ dominate when E$_{F}$ lies above the conduction band minimum (positive-bias stress), whereas H$^{+}$ is favored when E$_{F}$ lies below the valence band maximum (negative-bias stress) \cite{MFathi2025IRPS}. Under positive-bias stress, H$^{-}$ and H$^{0}$ therefore dominate.

To investigate H diffusion in amorphous sub-stoichiometric indium oxide under applied bias, H-doped a-InO was modeled. Single H with different effective charge states ("+", "0" and "-") were introduced near oxygen or indium atoms in a disordered 133-atom supercell. MD simulations at RT were then performed for 266 configurations containing a single H atom, and the nearest-neighbor environment of each H was tracked throughout the simulations. The results are summarized in Fig. 13. A large number of In–In hopping events is observed for H$^{0}$ (2,765 hops) and H$^{-}$ (7,742 hops) within a 2 pico-second (ps) simulation window. This hopping facilitates passivation of In–In pairs and the formation of shallow In–H–In donor states as shown in Fig. 12(b), increasing the channel carrier concentration during stress and leading to a strong negative $\Delta$V$_{TH}$. Elevated temperature further enhances H mobility, intensifying this effect. With prolonged FGA (60 min), partial crystallization occurs, generating additional c/a interfaces. These interfaces promote the formation of stable O–H bonds, which trap intrinsic H, suppress its motion during stress, and reduce the magnitude of the negative $\Delta$V$_{TH}$. The complete band-diagram-based mechanism is illustrated in Fig. 12(b).

Fig. 12(c) summarizes the NBTI trends across different FGA durations. Before FGA, the devices exhibit a negative $\Delta$V$_{TH}$, originating from negative electric-field-induced ionization of oxygen vacancies in the oxide channel, as illustrated by the reaction scheme in Fig. 12(d). The released electrons increase channel conductivity, resulting in a negative $\Delta$V$_{TH}$. As the FGA duration increases, $\Delta$V$_{TH}$ gradually shifts in the positive direction, with the effect becoming more pronounced at higher operating temperatures.

This behavior arises because, under negative-bias stress, the H$^{+}$ charge state dominates. H$^{+}$ preferentially forms strong O–H bonds, consistent with the simulation results in Fig. 13. With longer FGA, partial crystallization produces more c/a interfaces that serve as favorable sites for O–H bond formation. Once H is immobilized at these interfaces, fewer In–In pairs are passivated, leaving localized electron traps in the channel. Consequently, the effective carrier density decreases, leading to a positive $\Delta$V$_{TH}$.

\section{Conclusion}
In this work, we demonstrate a robust hybrid capping strategy that preserves OS-TFT performance after 400$^\circ$C post-processing FGA for up to 60 min. We further elucidate the evolution of electrical characteristics during FGA, including the emergence and suppression of the transfer-curve “kink” and the associated BTI behavior. Through DFT modeling, a two-stage defect-dynamics process is identified, consisting of initial film densification followed by partial crystallization. Finally, by integrating experimental results with modeling insights, we establish a comprehensive BTI mechanism that captures the bias- and annealing-dependent behavior of OS-TFTs. These findings provide a clear pathway for reliable integration of OS-TFTs into advanced BEOL process flows.

\bibliographystyle{IEEEtran}

\bibliography{ref.bib}

\end{document}